\documentclass{article}

\usepackage[final]{tackling_climate_workshop_style}
\usepackage[title]{appendix}

\usepackage[utf8]{inputenc} % allow utf-8 input
\usepackage[T1]{fontenc}    % use 8-bit T1 fonts
\usepackage{hyperref}       % hyperlinks
\usepackage{url}            % simple URL typesetting
\usepackage{booktabs}       % professional-quality tables
\usepackage{amsfonts}       % blackboard math symbols
\usepackage{nicefrac}       % compact symbols for 1/2, etc.
\usepackage{microtype}      % microtypography

\usepackage{graphicx}

\title{Short-range forecasts of global precipitation using deep learning-augmented numerical weather prediction}

\author{%
  Manmeet Singh \\
   Jackson School of Geosciences\\
  The University of Texas at Austin, \\
   Austin, TX 78712 \\
  \texttt{manmeet.singh@utexas.edu} \\
   \And
   Vaisakh SB \\
  Indian Institute of Tropical Meteorology\\
  Ministry of Earth Sciences\\
   Pune, India 411008 \\
  \texttt{vaisakh.sb@tropmet.res.in} \\
   \AND
   Nachiketa Acharya \\
  CIRES, University of Colorado Boulder\\
  NOAA/Physical Sciences Laboratory\\
   Boulder, CO 80309 \\
  \texttt{dr.nachiketaacharya@gmail.com} \\
   \And
   Aditya Grover \\
  Department of Computer Science\\
  University of California, Los Angeles\\
   Los Angeles, CA 90095 \\
  \texttt{adityag@cs.ucla.edu} \\
   \And
   Suryachandra A. Rao \\
Indian Institute of Tropical Meteorology\\
  Ministry of Earth Sciences\\
   Pune, India 411008 \\
  \texttt{surya@tropmet.res.in} \\
   \And
    Bipin Kumar \\
Indian Institute of Tropical Meteorology\\
  Ministry of Earth Sciences\\
   Pune, India 411008 \\
  \texttt{bipink@tropmet.res.in} \\
   \And
   Zong-Liang Yang \\
  Jackson School of Geosciences\\
  The University of Texas at Austin, \\
   Austin, TX 78712 \\
  \texttt{liang@jsg.utexas.edu} \\
  \And
     Dev Niyogi \\
  Jackson School of Geosciences\\
  The University of Texas at Austin, \\
   Austin, TX 78712 \\
  \texttt{dev.niyogi@jsg.utexas.edu} \\
}

\begin{document}

\maketitle

\begin{abstract}
Precipitation drives the hydroclimate of Earth and its spatiotemporal changes on a day to day basis have one of the most notable socioeconomic impacts. The success of numerical weather prediction (NWP) is measured by the improvement of forecasts for various physical fields such as temperature and pressure. Large biases however exist in the precipitation predictions. Pure deep learning based approaches lack the advancements acheived by NWP in the past two to three decades. Hybrid methodology using NWP outputs as inputs to the deep learning based refinement tool offer an attractive means taking advantage of both NWP and state of the art deep learning algorithms. Augmenting the output from a well-known NWP model: Coupled Forecast System ver.2 (CFSv2) with deep learning for the first time, we demonstrate a hybrid model capability (\textbf{DeepNWP}) which shows substantial skill improvements for short-range global precipitation at 1-, 2- and 3-days lead time. To achieve this hybridization, we address the sphericity of the global data by using modified DLWP-CS architecture which transforms all the fields to cubed-sphere projection. The dynamical model outputs corresponding to precipitation and surface temperature are ingested to a UNET for predicting the target ground truth precipitation. While the dynamical model CFSv2 shows a bias in the range of +5 to +7 mm/day over land, the multivariate deep learning model reduces it to -1 to +1 mm/day over global land areas. We validate the results by taking examples from Hurricane Katrina in 2005, Hurricane Ivan in 2004, Central European floods in 2010, China floods in 2010, India floods in 2005 and the Myanmar cyclone Nargis in 2008.  
\end{abstract}

\section{Introduction}
\label{sec:intro}
Precipitation forms an essential component of the hydrological cycle for the planet. The precipitation falling on land and oceans provides sustenance to the living beings and its accurate predictions can lead to tremendous societal benefits. At short-range time scales, i.e. of the order one to three days, the information of rain occurrence is helpful for various stakeholders. For example, ahead in time knowledge of precipitation can help better mitigate or manage the impacts of hurricanes, cyclones and floods. At present, most of the weather forecasting in the world is performed by the models which are known as dynamical models or numerical weather prediction (NWP) models \cite{bauer2015quiet}. Past few decades have seen improvement in the skill of numerical weather predictions, particularly that of temperature and winds. However, when it comes to physical fields such as precipitation and soil moisture, there are large biases in the outputs of these models \cite{schmidt2010real}. Deep learning can be effectively used as a tool to correct the biases in NWP models \cite{reichstein2019deep}. The ability of deep learning to unwrap nonlinear patterns in the data coupled with its success in the spatial tasks within the computer vision community makes it specially attractive for this task \cite{lecun2015deep}. There have been attempts to perform data-driven weather forecasting by various researchers in the past \cite{kashinath2021physics, rolnick2019tackling}. They were however limited by (i) performing training on linear fields relative to precipitation such as geopotential height at 500 hPa; thus not effectively addressing the physical field of direct interest to the humans \cite{bihlo2021generative, bihlo2022physics, chattopadhyay2020data, rasp2020weatherbench, scher2018toward, scher2019weather, weyn2019can, weyn2020improving}, (ii) their ability to address sphericity of the global data over Earth \cite{arcomano2020machine, bihlo2021generative, chattopadhyay2020data, han2020moist, mooers2021assessing, rasp2020weatherbench, rasp2018neural, rasp2018neural, scher2018toward, scher2019weather, shi2017deep, wang2020towards, yuval2020stable, mouatadid2021learned}, (iii) performing their study only over a limited region; thus neglecting the global teleconnections \cite{arcomano2020machine, bihlo2021generative, rasp2018neural, shi2017deep, mouatadid2021learned}, (iv) their framework not utilizing and building upwards on the advancements in NWP; and rather trying to achieve the results similar to NWP and not utlizing the advancements in NWP \& deep learning to develop hybrid forecasting framework to enhance the skill of the existing systems \cite{arcomano2020machine, bihlo2021generative, bihlo2022physics, chattopadhyay2020data, mooers2021assessing, rasp2020weatherbench, shi2017deep, weyn2019can, weyn2020improving, yuval2020stable}.
\par 
\begin{figure}
  \centering
  \includegraphics[width=1.0\linewidth]{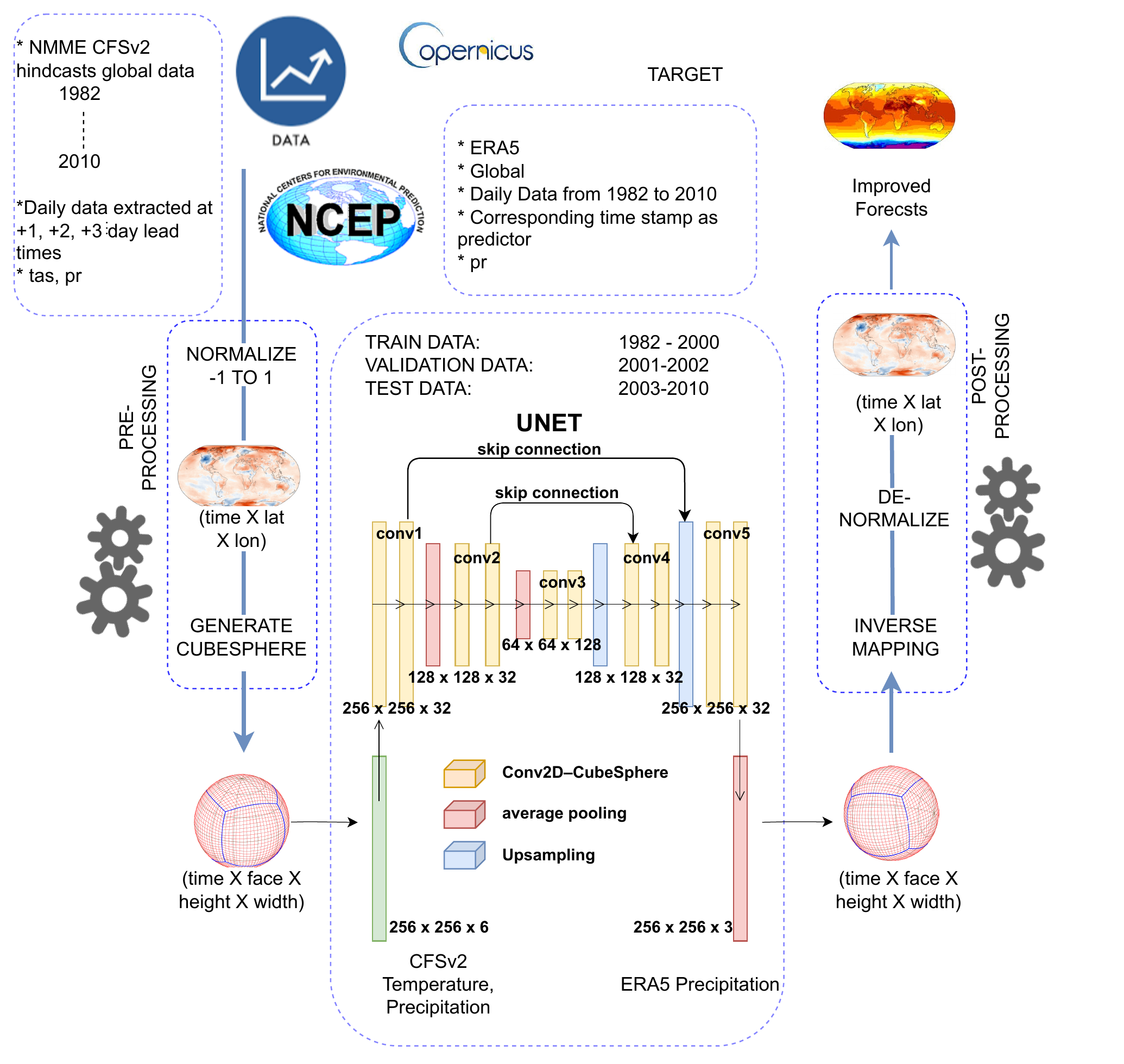}

  \caption{Schematic of the deep learning-augmented numerical weather prediction.}
  \label{fig:short}
\end{figure}

\par

\par 
\subsection{Our contributions}
Rather than using deep learning directly to forecast precipitation, we use a hybrid approach combining it with the traditional NWP. In particular, we consider CFSv2 which is a model used for operational weather prediction in the United States, India and other countries. We propose to correct its biases by training a deep neural network that projects its outputs to ERA5 reanalysis, which is a merged product of ground and space-based observations and data-assimilation available from European Centre for Medium-Range Weather Forecasts is used as a target. ERA5 reanalysis precipitation has shown high fidelity in representing preciptiation in the recent times \cite{gleixner2020did}. Thus, we use the outputs of precipitation and temperature from CFSv2 model and map them to the ERA5 reanalysis precipitation. The architecture of deep learning based computer vision algorithm used in modified DLWP-CS \cite{singh2021deep}.

Our NWP augmented deep learning product shows substantial improvements in the representation of precipitation. On an average, we are able to reduce the bias in precipitation from beyond +10 mm/day to within +- 1 mm/day. The results are consistent for different lead days in the short range time scale (+1, +2 and +3 days). When segregated based on the different seasons, the global mean bias values show improvevement upwards of 10x by introducing deep learning in the numerical weather prediction system. We validate our results with specific use cases comprising of heavy to extreme precipitation from the test data. The use cases corroborate the findings from the global mean bias and show even more substantial improvements, the maximum being for India flood 2005 wherein the bias in NWP decreased from +17.321 mm/day to +0.003 mm/day for day 1 lead forecast when deep learning is added to the system. Our hybrid model can be used for mitigating the impacts of various natural disasters triggered by precipitation and effective management of water resources.  

\section{Data and methodology}

The input dataset comprises of precursors, viz, CFSv2 generated precipitation and surface air temeprature outputs. Precipitation and temperature outputs of CFSv2 are downloaded from North American Multi-Model Ensemble (NMME) for developing the input training dataset. ERA5 reanalysis \cite{hersbach2020era5} is used as the ground truth representing the real-world precipitation. We obtain the precursors from the CFSv2 model outputs and the target extracted from ERA5 reanalysis. The CFSv2 output is available as 6-hourly average for a lead time of upto +10 months while the ERA5 reanalysis is available an hourly product. Daily aggregates are first generated for the input (CFSv2) and target (ERA5) by summing the sub-daily datasets. Then, the time slices which represent 1, 2 and 3 days lead in CFSv2 are extracted. For the labels from ERA5, the data is extracted such that it corresponds to the forecasted 1, 2 and 3 days from CFSv2 in reality. Data from 1982 to 2000 is used for training, 2000-2001 for validation and 2003 to 2010 for testing. The maximum and minimum values of the training data fields are then computed for normalizing the data. Normalization scales the different variables to the range -1 to +1. We call our model as \textbf{DeepNWP} to signify the hybrid nature of deep learning augmented numerical weather prediction. The methodology used is shown as a schematic in figure \ref{fig:short}

\par
\begin{figure*}
  \centering
  \includegraphics[width=0.8\linewidth]{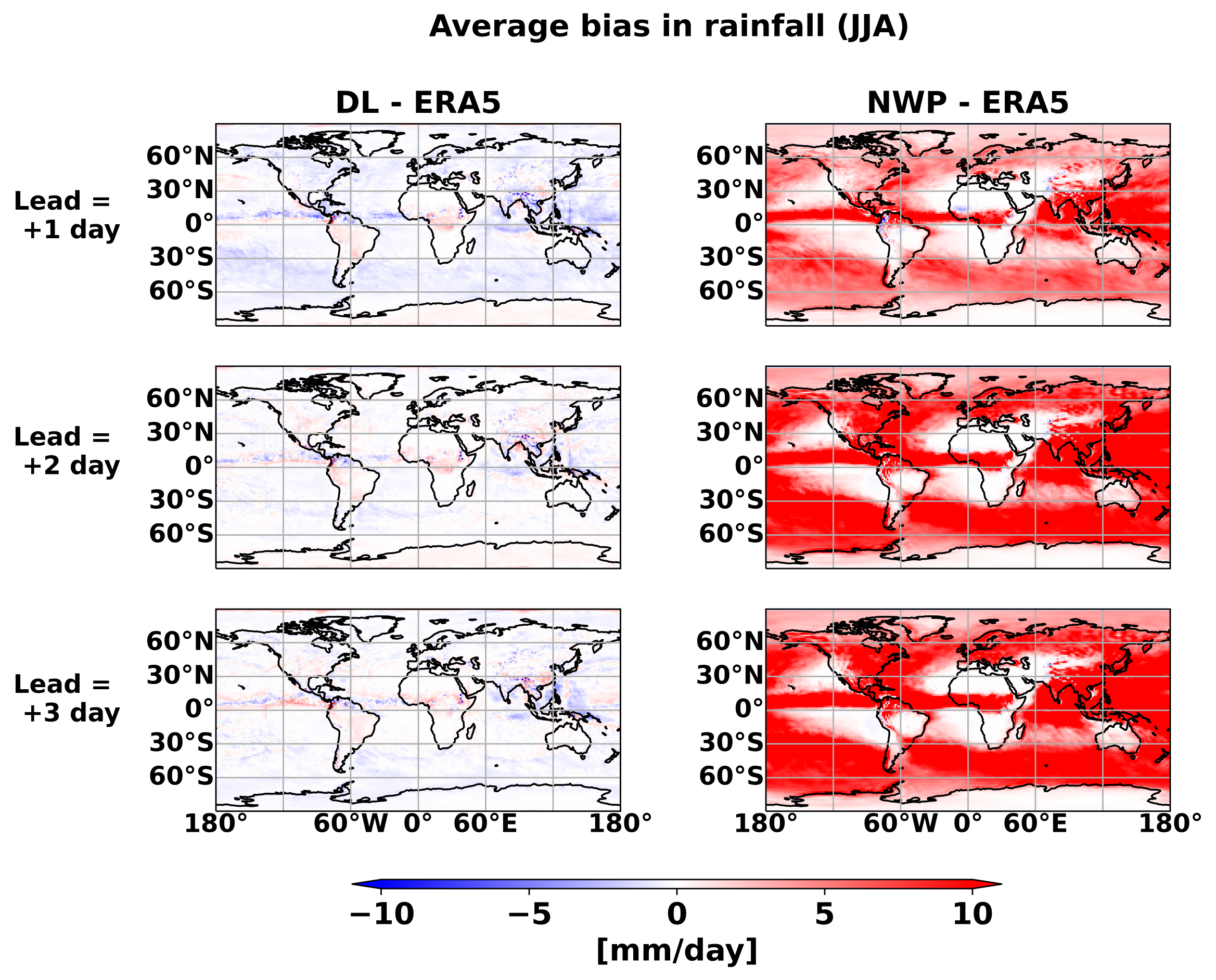}

  \caption{Average bias in precipitation for June to August season for the deep learning augmented NWP model (DL) and NWP model relative to ERA5 reanalysis. The rows represent the bias at lead time = 1, 2 and 3 days. }
  \label{fig:bias}
\end{figure*}

\begin{table*}
%  \centering
  \begin{tabular}{||c||c|c||c|c||c|c||}
\hline
\hline
\multicolumn{1}{|c|}{Season}
  & \multicolumn{2}{|c||}{Lead = 1 day (mm/day)} & \multicolumn{2}{|c||}{Lead = 2 day (mm/day)} & \multicolumn{2}{|c||}{Lead = 3 day (mm/day)}\\	
% \hline
\cline{2-7}
  &  DL - ERA5 & CFSv2 - ERA5 &  DL - ERA5 & CFSv2 - ERA5 &  DL - ERA5 & CFSv2 - ERA5\\
 \hline
\hline

DJF & \textbf{-0.3} & 3.827 & \textbf{0.022} & 7.66 & \textbf{-0.158} & 7.657\\
MAM & \textbf{-0.282} & 3.834 & \textbf{0.032} & 7.811 & \textbf{-0.11} & 7.89\\
JJA & \textbf{-0.334} & 3.97 & \textbf{-0.02} & 8.102 & \textbf{-0.115} & 8.239\\
SON & \textbf{-0.299} & 3.954 & \textbf{0} & 7.95 & \textbf{-0.148} & 7.972\\

\hline
\hline
  \end{tabular}
  \caption{Performance of the deep learning augmented numerical weather prediction system DeepNWP (bold) versus CFSv2 alone. The table shows global average bias/error in simulating precipitation by the hybrid deep learning and CFSv2 system versus CFSv2 alone. DJF (December to February), MAM (March to May), JJA (June to August) and SON (September to November) represent the different months of an year. The performance is shown for the entire test period from the year 2003 to 2010.}
  \label{tab:tab1}
\end{table*}

\section{Results}
Figure \ref{fig:bias} shows the performance measured by average bias or error in deep learning augmented NWP model (DL) and NWP model alone relative to ERA5 reanalysis during June to August season. The first column corresponds to DL bias and second column represents the NWP model bias. The average bias or error is computed as a mean of the difference between the model (DL or NWP) and ERA5 reanalysis for all the samples in the test dataset. We also compute the plots similar to figure \ref{fig:bias} for other seasons, viz, December to February, March to May and September to November (figures not shown). From the average seasonal bias figures, it can be noted that deep learning augmented NWP model substantially reduces the bias in precipitation relative to the NWP (CFSv2) model. While the mean bias over important land regions lies within the range +5 to +7 mm/day in NWP (CFSv2) (similar to the results of \cite{mukhopadhyay2019performance}), it falls to within the range -1 to +1 mm/day using deep learning. Thus, in a mean sense, deep learning augmented CFSv2 model improves the performance of precipitation forecasts by upto 4-5x. We can note the enhanced performance until the computed +3 days lead time for deep learning augmented NWP model. Table \ref{tab:tab1} shows the bias (model - ground truth) in global mean precipitation at different lead times (1, 2 and 3 days). Substantially enhanced skill of global precipitation forecasts upwards of 10x is noted for the all the seasons.

\section{Conclusions}

In this study, we develop a hybrid deep learning augmented numerical weather prediction (DeepNWP) model for generating global precipitation forecasts at short-range (1, 2 and 3 days) time scales. While previous studies have noted the possibility of a hybrid deep learning augmented NWP model and also suggesting an end-to-end deep learning based system \cite{schultz2021can, weyn2019can}, ours is a first attempt in actually developing such a system. All the existing implementations using deep learning for weather prediction have attempted for relatively simplistic fields such as geopotential height, while we actually attempt to improve global precipitation forecasts. We ensure that the sphericity of the global data over Earth is considered and find vast improvements in deep learning augmented NWP forecasts relative to the NWP alone. In future, we would use more precursors of precipitation from the NWP model for generating deep learning augmented weather predictions. Further, hyperparameter tuning and neural architecture search would be performed for performance enhancement.

{\small
\bibliographystyle{plainnat}
\bibliography{egbib}

\begin{thebibliography}{36}
\providecommand{\natexlab}[1]{#1}
\providecommand{\url}[1]{\texttt{#1}}
\expandafter\ifx\csname urlstyle\endcsname\relax
  \providecommand{\doi}[1]{doi: #1}\else
  \providecommand{\doi}{doi: \begingroup \urlstyle{rm}\Url}\fi

\bibitem[Arcomano et~al.(2020)Arcomano, Szunyogh, Pathak, Wikner, Hunt, and
  Ott]{arcomano2020machine}
Troy Arcomano, Istvan Szunyogh, Jaideep Pathak, Alexander Wikner, Brian~R Hunt,
  and Edward Ott.
\newblock A machine learning-based global atmospheric forecast model.
\newblock \emph{Geophysical Research Letters}, 47\penalty0 (9):\penalty0
  e2020GL087776, 2020.

\bibitem[Bauer et~al.(2015)Bauer, Thorpe, and Brunet]{bauer2015quiet}
Peter Bauer, Alan Thorpe, and Gilbert Brunet.
\newblock The quiet revolution of numerical weather prediction.
\newblock \emph{Nature}, 525\penalty0 (7567):\penalty0 47--55, 2015.

\bibitem[Bihlo(2021)]{bihlo2021generative}
Alex Bihlo.
\newblock A generative adversarial network approach to (ensemble) weather
  prediction.
\newblock \emph{Neural Networks}, 139:\penalty0 1--16, 2021.

\bibitem[Bihlo and Popovych(2022)]{bihlo2022physics}
Alex Bihlo and Roman~O Popovych.
\newblock Physics-informed neural networks for the shallow-water equations on
  the sphere.
\newblock \emph{Journal of Computational Physics}, page 111024, 2022.

\bibitem[Chattopadhyay et~al.(2020)Chattopadhyay, Hassanzadeh, and
  Subramanian]{chattopadhyay2020data}
Ashesh Chattopadhyay, Pedram Hassanzadeh, and Devika Subramanian.
\newblock Data-driven predictions of a multiscale lorenz 96 chaotic system
  using machine-learning methods: reservoir computing, artificial neural
  network, and long short-term memory network.
\newblock \emph{Nonlinear Processes in Geophysics}, 27\penalty0 (3):\penalty0
  373--389, 2020.

\bibitem[Clough et~al.(2005)Clough, Shephard, Mlawer, Delamere, Iacono,
  Cady-Pereira, Boukabara, and Brown]{clough2005atmospheric}
SA~Clough, MW~Shephard, EJ~Mlawer, JS~Delamere, MJ~Iacono, K~Cady-Pereira,
  S~Boukabara, and PD~Brown.
\newblock Atmospheric radiative transfer modeling: A summary of the aer codes.
\newblock \emph{Journal of Quantitative Spectroscopy and Radiative Transfer},
  91\penalty0 (2):\penalty0 233--244, 2005.

\bibitem[Ek et~al.(2003)Ek, Mitchell, Lin, Rogers, Grunmann, Koren, Gayno, and
  Tarpley]{ek2003implementation}
MB~Ek, KE~Mitchell, Ying Lin, Eric Rogers, Pablo Grunmann, Victor Koren, George
  Gayno, and JD~Tarpley.
\newblock Implementation of noah land surface model advances in the national
  centers for environmental prediction operational mesoscale eta model.
\newblock \emph{Journal of Geophysical Research: Atmospheres}, 108\penalty0
  (D22), 2003.

\bibitem[Gleixner et~al.(2020)Gleixner, Demissie, and Diro]{gleixner2020did}
Stephanie Gleixner, Teferi Demissie, and Gulilat~Tefera Diro.
\newblock Did era5 improve temperature and precipitation reanalysis over east
  africa?
\newblock \emph{Atmosphere}, 11\penalty0 (9):\penalty0 996, 2020.

\bibitem[Griffies et~al.(2004)Griffies, Harrison, Pacanowski, and
  Rosati]{griffies2004technical}
Stephen~M Griffies, Matthew~J Harrison, Ronald~C Pacanowski, and Anthony
  Rosati.
\newblock A technical guide to mom4.
\newblock \emph{GFDL Ocean Group Tech. Rep}, 5:\penalty0 342, 2004.

\bibitem[Han et~al.(2020)Han, Zhang, Huang, and Wang]{han2020moist}
Yilun Han, Guang~J Zhang, Xiaomeng Huang, and Yong Wang.
\newblock A moist physics parameterization based on deep learning.
\newblock \emph{Journal of Advances in Modeling Earth Systems}, 12\penalty0
  (9):\penalty0 e2020MS002076, 2020.

\bibitem[Hersbach et~al.(2020)Hersbach, Bell, Berrisford, Hirahara,
  Hor{\'a}nyi, Mu{\~n}oz-Sabater, Nicolas, Peubey, Radu, Schepers,
  et~al.]{hersbach2020era5}
Hans Hersbach, Bill Bell, Paul Berrisford, Shoji Hirahara, Andr{\'a}s
  Hor{\'a}nyi, Joaqu{\'\i}n Mu{\~n}oz-Sabater, Julien Nicolas, Carole Peubey,
  Raluca Radu, Dinand Schepers, et~al.
\newblock The era5 global reanalysis.
\newblock \emph{Quarterly Journal of the Royal Meteorological Society},
  146\penalty0 (730):\penalty0 1999--2049, 2020.

\bibitem[Iacono et~al.(2000)Iacono, Mlawer, Clough, and
  Morcrette]{iacono2000impact}
Michael~J Iacono, Eli~J Mlawer, Shepard~A Clough, and Jean-Jacques Morcrette.
\newblock Impact of an improved longwave radiation model, rrtm, on the energy
  budget and thermodynamic properties of the ncar community climate model,
  ccm3.
\newblock \emph{Journal of Geophysical Research: Atmospheres}, 105\penalty0
  (D11):\penalty0 14873--14890, 2000.

\bibitem[Kashinath et~al.(2021)Kashinath, Mustafa, Albert, Wu, Jiang,
  Esmaeilzadeh, Azizzadenesheli, Wang, Chattopadhyay, Singh,
  et~al.]{kashinath2021physics}
K~Kashinath, M~Mustafa, A~Albert, JL~Wu, C~Jiang, S~Esmaeilzadeh,
  K~Azizzadenesheli, R~Wang, A~Chattopadhyay, A~Singh, et~al.
\newblock Physics-informed machine learning: case studies for weather and
  climate modelling.
\newblock \emph{Philosophical Transactions of the Royal Society A},
  379\penalty0 (2194):\penalty0 20200093, 2021.

\bibitem[Kim and Arakawa(1995)]{kim1995improvement}
Young-Joon Kim and Akio Arakawa.
\newblock Improvement of orographic gravity wave parameterization using a
  mesoscale gravity wave model.
\newblock \emph{Journal of Atmospheric Sciences}, 52\penalty0 (11):\penalty0
  1875--1902, 1995.

\bibitem[LeCun et~al.(2015)LeCun, Bengio, and Hinton]{lecun2015deep}
Yann LeCun, Yoshua Bengio, and Geoffrey Hinton.
\newblock Deep learning.
\newblock \emph{nature}, 521\penalty0 (7553):\penalty0 436--444, 2015.

\bibitem[Lott and Miller(1997)]{lott1997new}
Fran{\c{c}}ois Lott and Martin~J Miller.
\newblock A new subgrid-scale orographic drag parametrization: Its formulation
  and testing.
\newblock \emph{Quarterly Journal of the Royal Meteorological Society},
  123\penalty0 (537):\penalty0 101--127, 1997.

\bibitem[Mooers et~al.(2021)Mooers, Pritchard, Beucler, Ott, Yacalis, Baldi,
  and Gentine]{mooers2021assessing}
Griffin Mooers, Michael Pritchard, Tom Beucler, Jordan Ott, Galen Yacalis,
  Pierre Baldi, and Pierre Gentine.
\newblock Assessing the potential of deep learning for emulating cloud
  superparameterization in climate models with real-geography boundary
  conditions.
\newblock \emph{Journal of Advances in Modeling Earth Systems}, 13\penalty0
  (5):\penalty0 e2020MS002385, 2021.

\bibitem[Mouatadid et~al.(2021)Mouatadid, Orenstein, Flaspohler, Oprescu,
  Cohen, Wang, Knight, Geogdzhayeva, Levang, Fraenkel,
  et~al.]{mouatadid2021learned}
Soukayna Mouatadid, Paulo Orenstein, Genevieve Flaspohler, Miruna Oprescu,
  Judah Cohen, Franklyn Wang, Sean Knight, Maria Geogdzhayeva, Sam Levang,
  Ernest Fraenkel, et~al.
\newblock Learned benchmarks for subseasonal forecasting.
\newblock \emph{arXiv preprint arXiv:2109.10399}, 2021.

\bibitem[Mukhopadhyay et~al.(2019)Mukhopadhyay, Prasad, Krishna, Deshpande,
  Ganai, Tirkey, Sarkar, Goswami, Johny, Roy,
  et~al.]{mukhopadhyay2019performance}
P~Mukhopadhyay, VS~Prasad, R~Krishna, Medha Deshpande, Malay Ganai, Snehlata
  Tirkey, Sahadat Sarkar, Tanmoy Goswami, CJ~Johny, Kumar Roy, et~al.
\newblock Performance of a very high-resolution global forecast system model
  (gfs t1534) at 12.5 km over the indian region during the 2016--2017 monsoon
  seasons.
\newblock \emph{Journal of Earth System Science}, 128\penalty0 (6):\penalty0
  1--18, 2019.

\bibitem[Rasp and Lerch(2018)]{rasp2018neural}
Stephan Rasp and Sebastian Lerch.
\newblock Neural networks for postprocessing ensemble weather forecasts.
\newblock \emph{Monthly Weather Review}, 146\penalty0 (11):\penalty0
  3885--3900, 2018.

\bibitem[Rasp et~al.(2020)Rasp, Dueben, Scher, Weyn, Mouatadid, and
  Thuerey]{rasp2020weatherbench}
Stephan Rasp, Peter~D Dueben, Sebastian Scher, Jonathan~A Weyn, Soukayna
  Mouatadid, and Nils Thuerey.
\newblock Weatherbench: a benchmark data set for data-driven weather
  forecasting.
\newblock \emph{Journal of Advances in Modeling Earth Systems}, 12\penalty0
  (11):\penalty0 e2020MS002203, 2020.

\bibitem[Reichstein et~al.(2019)Reichstein, Camps-Valls, Stevens, Jung,
  Denzler, Carvalhais, et~al.]{reichstein2019deep}
Markus Reichstein, Gustau Camps-Valls, Bjorn Stevens, Martin Jung, Joachim
  Denzler, Nuno Carvalhais, et~al.
\newblock Deep learning and process understanding for data-driven earth system
  science.
\newblock \emph{Nature}, 566\penalty0 (7743):\penalty0 195--204, 2019.

\bibitem[Rolnick et~al.(2019)Rolnick, Donti, Kaack, Kochanski, Lacoste,
  Sankaran, Ross, Milojevic-Dupont, Jaques, Waldman-Brown,
  et~al.]{rolnick2019tackling}
David Rolnick, Priya~L Donti, Lynn~H Kaack, Kelly Kochanski, Alexandre Lacoste,
  Kris Sankaran, Andrew~Slavin Ross, Nikola Milojevic-Dupont, Natasha Jaques,
  Anna Waldman-Brown, et~al.
\newblock Tackling climate change with machine learning.
\newblock \emph{arXiv preprint arXiv:1906.05433}, 2019.

\bibitem[Saha et~al.(2014)Saha, Moorthi, Wu, Wang, Nadiga, Tripp, Behringer,
  Hou, Chuang, Iredell, et~al.]{saha2014ncep}
Suranjana Saha, Shrinivas Moorthi, Xingren Wu, Jiande Wang, Sudhir Nadiga,
  Patrick Tripp, David Behringer, Yu-Tai Hou, Hui-ya Chuang, Mark Iredell,
  et~al.
\newblock The ncep climate forecast system version 2.
\newblock \emph{Journal of climate}, 27\penalty0 (6):\penalty0 2185--2208,
  2014.

\bibitem[Scher(2018)]{scher2018toward}
Sebastian Scher.
\newblock Toward data-driven weather and climate forecasting: Approximating a
  simple general circulation model with deep learning.
\newblock \emph{Geophysical Research Letters}, 45\penalty0 (22):\penalty0
  12--616, 2018.

\bibitem[Scher and Messori(2019)]{scher2019weather}
Sebastian Scher and Gabriele Messori.
\newblock Weather and climate forecasting with neural networks: using general
  circulation models (gcms) with different complexity as a study ground.
\newblock \emph{Geoscientific Model Development}, 12\penalty0 (7):\penalty0
  2797--2809, 2019.

\bibitem[Schmidt(2010)]{schmidt2010real}
Gavin Schmidt.
\newblock The real holes in climate science.
\newblock \emph{Nature}, 463:\penalty0 21, 2010.

\bibitem[Schultz et~al.(2021)Schultz, Betancourt, Gong, Kleinert, Langguth,
  Leufen, Mozaffari, and Stadtler]{schultz2021can}
MG~Schultz, Clara Betancourt, Bing Gong, Felix Kleinert, Michael Langguth,
  LH~Leufen, Amirpasha Mozaffari, and Scarlet Stadtler.
\newblock Can deep learning beat numerical weather prediction?
\newblock \emph{Philosophical Transactions of the Royal Society A},
  379\penalty0 (2194):\penalty0 20200097, 2021.

\bibitem[Shi et~al.(2017)Shi, Gao, Lausen, Wang, Yeung, Wong, and
  Woo]{shi2017deep}
Xingjian Shi, Zhihan Gao, Leonard Lausen, Hao Wang, Dit-Yan Yeung, Wai-kin
  Wong, and Wang-chun Woo.
\newblock Deep learning for precipitation nowcasting: A benchmark and a new
  model.
\newblock \emph{Advances in neural information processing systems}, 30, 2017.

\bibitem[Singh et~al.(2021)Singh, Kumar, Rao, Gill, Chattopadhyay, Nanjundiah,
  and Niyogi]{singh2021deep}
Manmeet Singh, Bipin Kumar, Suryachandra Rao, Sukhpal~Singh Gill, Rajib
  Chattopadhyay, Ravi~S Nanjundiah, and Dev Niyogi.
\newblock Deep learning for improved global precipitation in numerical weather
  prediction systems.
\newblock \emph{arXiv preprint arXiv:2106.12045}, 2021.

\bibitem[Wang et~al.(2020)Wang, Kashinath, Mustafa, Albert, and
  Yu]{wang2020towards}
Rui Wang, Karthik Kashinath, Mustafa Mustafa, Adrian Albert, and Rose Yu.
\newblock Towards physics-informed deep learning for turbulent flow prediction.
\newblock In \emph{Proceedings of the 26th ACM SIGKDD International Conference
  on Knowledge Discovery \& Data Mining}, pages 1457--1466, 2020.

\bibitem[Weyn et~al.(2019)Weyn, Durran, and Caruana]{weyn2019can}
Jonathan~A Weyn, Dale~R Durran, and Rich Caruana.
\newblock Can machines learn to predict weather? using deep learning to predict
  gridded 500-hpa geopotential height from historical weather data.
\newblock \emph{Journal of Advances in Modeling Earth Systems}, 11\penalty0
  (8):\penalty0 2680--2693, 2019.

\bibitem[Weyn et~al.(2020)Weyn, Durran, and Caruana]{weyn2020improving}
Jonathan~A Weyn, Dale~R Durran, and Rich Caruana.
\newblock Improving data-driven global weather prediction using deep
  convolutional neural networks on a cubed sphere.
\newblock \emph{Journal of Advances in Modeling Earth Systems}, 12\penalty0
  (9):\penalty0 e2020MS002109, 2020.

\bibitem[Winton(2000)]{winton2000reformulated}
Michael Winton.
\newblock A reformulated three-layer sea ice model.
\newblock \emph{Journal of atmospheric and oceanic technology}, 17\penalty0
  (4):\penalty0 525--531, 2000.

\bibitem[Wu et~al.(1997)Wu, Simmonds, and Budd]{wu1997modeling}
Xingren Wu, Ian Simmonds, and WF~Budd.
\newblock Modeling of antarctic sea ice in a general circulation model.
\newblock \emph{Journal of Climate}, 10\penalty0 (4):\penalty0 593--609, 1997.

\bibitem[Yuval and O’Gorman(2020)]{yuval2020stable}
Janni Yuval and Paul~A O’Gorman.
\newblock Stable machine-learning parameterization of subgrid processes for
  climate modeling at a range of resolutions.
\newblock \emph{Nature communications}, 11\penalty0 (1):\penalty0 1--10, 2020.

\end{thebibliography}
}

% \section*{References}

% References follow the acknowledgments. Use unnumbered first-level heading for
% the references. Any choice of citation style is acceptable as long as you are
% consistent. It is permissible to reduce the font size to \verb+small+ (9 point)
% when listing the references.
% {\bf Note that the Reference section does not count towards the pages of content that are allowed; 4 pages for Papers track and 3 pages for Proposals track.}
% \medskip

% \small

% [1] Alexander, J.A.\ \& Mozer, M.C.\ (1995) Template-based algorithms for
% connectionist rule extraction. In G.\ Tesauro, D.S.\ Touretzky and T.K.\ Leen
% (eds.), {\it Advances in Neural Information Processing Systems 7},
% pp.\ 609--616. Cambridge, MA: MIT Press.

% [2] Bower, J.M.\ \& Beeman, D.\ (1995) {\it The Book of GENESIS: Exploring
%   Realistic Neural Models with the GEneral NEural SImulation System.}  New York:
% TELOS/Springer--Verlag.

% [3] Hasselmo, M.E., Schnell, E.\ \& Barkai, E.\ (1995) Dynamics of learning and
% recall at excitatory recurrent synapses and cholinergic modulation in rat
% hippocampal region CA3. {\it Journal of Neuroscience} {\bf 15}(7):5249-5262.

\begin{appendices}
\section{Training details}
The training is performed on a single NVIDIA A100 GPU with 40 GB GPU RAM. Since the size of data is huge for I/O, loading on GPU and training, efficient use of TensorFlow dataloaders is made. The total number of training samples is 5528 corresponding to the data from 1982 to 2000 with a batch size of 8. The batch size is chosen to ensure that the heavy data can be trained on the A100 GPU by iterating on different choices. We use the adam optimizer,  learning rate of 0.0001 and mean squared error as loss function. Since I/O was a big overhead for the training, it is performed in loop by training each year in one go and then improving the model for years that come after. We call it incremental training and to prevent overfitting, early stopping is used.
 \\
Incorporating the effects of sphericity while also using a computer vision based deep learning model, the latitude-longitude dataset is transformed to the cubed sphere mapping. We use the implementation of tempest remap to perform the cubed sphere transformation. Cubed sphere is typically a cubical sphere with 6 faces. Two of these faces represent the polar regions while four faces are over the tropics. The resolution of each face of the cubed sphere that is used in this study is 256 x 256. The choice of the cubed sphere resolution is determined by transforming from a regular latitude-longitude grid to cubed sphere and then back to the regular latitude longitude grid. Errors induced by the cubed sphere by a forward and backward pass are computed and the optimal cubed resolution of 256 x 256 is selected. \\
We use modified DLWP-CS \cite{singh2021deep} which is evolved from DLWP-CS \cite{weyn2020improving} for training the deep learning model. Modified DLWP-CS use a 2D UNET to perform an image to image regression for the cubed sphere data to map from one or multiple variables to uni or multivariate target. The core of modified DLWP-CS uses two 2D UNETs, one for the tropics and other for the poles. They are coupled to each other in such a way that the data is exchanged while ensuring the same input and output matrix size during convolution. Typically, in computer vision applications, a padding of zeros is used. However, a global spherical data doesn't have any real boundaries, so neighbouring edge of a cubed sphere face is used for padding. This data exchange process ensures that the important global linkages such as those to and from El Nino Southern Oscillation, North Atlantic Oscillation and Pacific Decadal Oscillation are considered by the data-driven approach. The different input variables from CFSv2 model; viz, precipitation and temperature are fed as channels to the UNET from modified DLWP-CS. The target is a single channel ERA5 precipitation in cubed sphere projection.
\section{Preliminaries}
Decades of scientific research into improving the global precipitation forecasts has led to the state of the art NWP systems which are behind the operational weather prediction worldwide. Although, NWP has been a success, problems remain in fields such as precipitation and soil moisture showing high bias in their representation in the NWP models. The NWP systems are basically solutions of governing equations of the atmosphere ocean and land which are coupled to each other. These partial differential equations are solved by assuming Earth to be a sphere and initializing from appropriate initial and boundary conditions. The equations are solved for prognostic fields such as pressure, temperature, winds and tracers such as water vapor and ozone. Precipitation and soil moisture are diagnostic fields which are derived by emperical relationships also known as parameterization schemes taking in the prognostic fields as input. These parameterization schemes are challenging to build and improve in the NWP systems and the problems with diagnostic fields are mostly attributed to them. CFSv2 is a model which is used operationally for weather forecasting by different nodal centers around the world. 

Recently, studies \cite{kashinath2021physics, rolnick2019tackling} have shown the capability of deep learning to emulate weather systems purely from a data-driven approach. However, these methods often lack the capability to deliver a global product which can be directly used for societal applications. More on the limitations of these studies has been discussed in section 1.1. Applications \cite{weyn2019can, weyn2020improving} have attempted to solve various challenges in pure data-driven models such as addressing the sphericity of Earth \cite{weyn2020improving} and using global datasets. However, they \cite{weyn2019can, weyn2020improving} lack the strength of a unified deep learning NWP system. We use ERA5 reanalysis global precipitation as the ground truth target. ERA5 reanalysis is a merged product of satellite, gauge-based ground observations, ocean buoys and other instrumental data combined with model outputs to generate a global product. It has shown fidelity to represent precipitation similar to the observations.

\section{Climate Forecast system version 2 (CFSv2)}
CFSv2 is a spectral model in which atmosphere, ocean and land compoents are coupled to each other. Within the CFSv2, the atmospheric model is known as the Global Forecast System (GFS). The atmospheric component of CFSv2, i.e. GFS has a spectral resolution has a spectral resolution of T126 with 64 vertical levels \cite{saha2014ncep}. Ocean model of CFSv2 is the Modular Ocean Model developed by the Geophysical Fluid Dynamics Laboratory (GFDL) \cite{griffies2004technical}. Arakawa-Schubert scheme is used for convective parameterization with momentum mixing. CFSv2 incorporates the effects such as orographic gravity wave drag and mountain blockage \cite{kim1995improvement,lott1997new}. Rapid radiative transfer model is used for the radiation computations in the atmosphere \cite{iacono2000impact, clough2005atmospheric}. A four-layer land-surface model known as the NOAH LSM \cite{ek2003implementation} and a dynamical two-layer sea ice model \cite{wu1997modeling,winton2000reformulated} are coupled to the atmosphere and ocean components of CFSv2. CFSv2 hindcast simulations from 1982 to 2010 are downloaded. The hindcast model simulations are initialized at an interval of 5 days for the years 1982-2010. A total number of 4 ensemble members are generated for each intialization day of the model. We use all the available ensembles for training. 
\section{ERA5 reanalysis}
ERA5 dataset is provided by the European Center for Medium-Range Weather Forecasts (ECMWF) and has been available since 2019. It consists of meteorological fields at hourly temporal intervals and a global spatial resolution of 0.25 degrees. The data is available from the year 1979 which was one of the reason for selecting ERA5 reanalysis as the ground truth. Other global datasets such as GPCC and GPCP are either available since the satellite era  (1996 onwards) precipitation products became available or are only over land. Recent studies have shown that ERA5 reanalysis precipitation provides a reasonable representation of the precipitation \cite{gleixner2020did}.
\section{Case studies}
We validate the skill of \textbf{DeepNWP} for specific examples viz, (i) Hurricane Katrina, (ii) Hurricane Ivan, (iii) Cyclone Nargis, (iv) Europe flood 2010, (v) China flood 2010 and (vi) India flood 2005. Spatial maps of absolute values corresponding to ERA5 (similar to observations), deep learning augmented CFSv2 and CFSv2 are shown at the lead time = 1, 2 and 3 days. The lead times are with reference to the starting date of CFSv2 mode, i.e. since the time CFSv2 gets initialized. In addition, the figures show difference in model
predicted precipitation relative to ERA5 for the three considered lead times. Considering the individual hurricane, cyclone and flooding events, it can be seen that CFSv2 has a much larger bias in the tropics relative to mid latitudes (figures \ref{fig:ivan}, \ref{fig:europe}, \ref{fig:china}, \ref{fig:india}, \ref{fig:nargis}). CFSv2 has a wet bias for the different events, even exceeding +20 mm/day over different regions across the selected cases. DeepNWP reduces the bias to within the range -1 to +1 mm/day except for cyclone Nargis. Cyclone Nargis shows humongous bias in CFSv2 from +20 to +47 mm/day while the deep learning model reduces these errors from -5.37 to +2.34 mm/day for the three lead times considered. A summary of these statistics is provided in the table \ref{tab:example}.

Following are some case studies from the test data period corresponding to the years 2003-2010 showing the superior performance of deep learning augmented NWP:

\begin{figure*}
  \centering
  \includegraphics[width=0.8\linewidth]{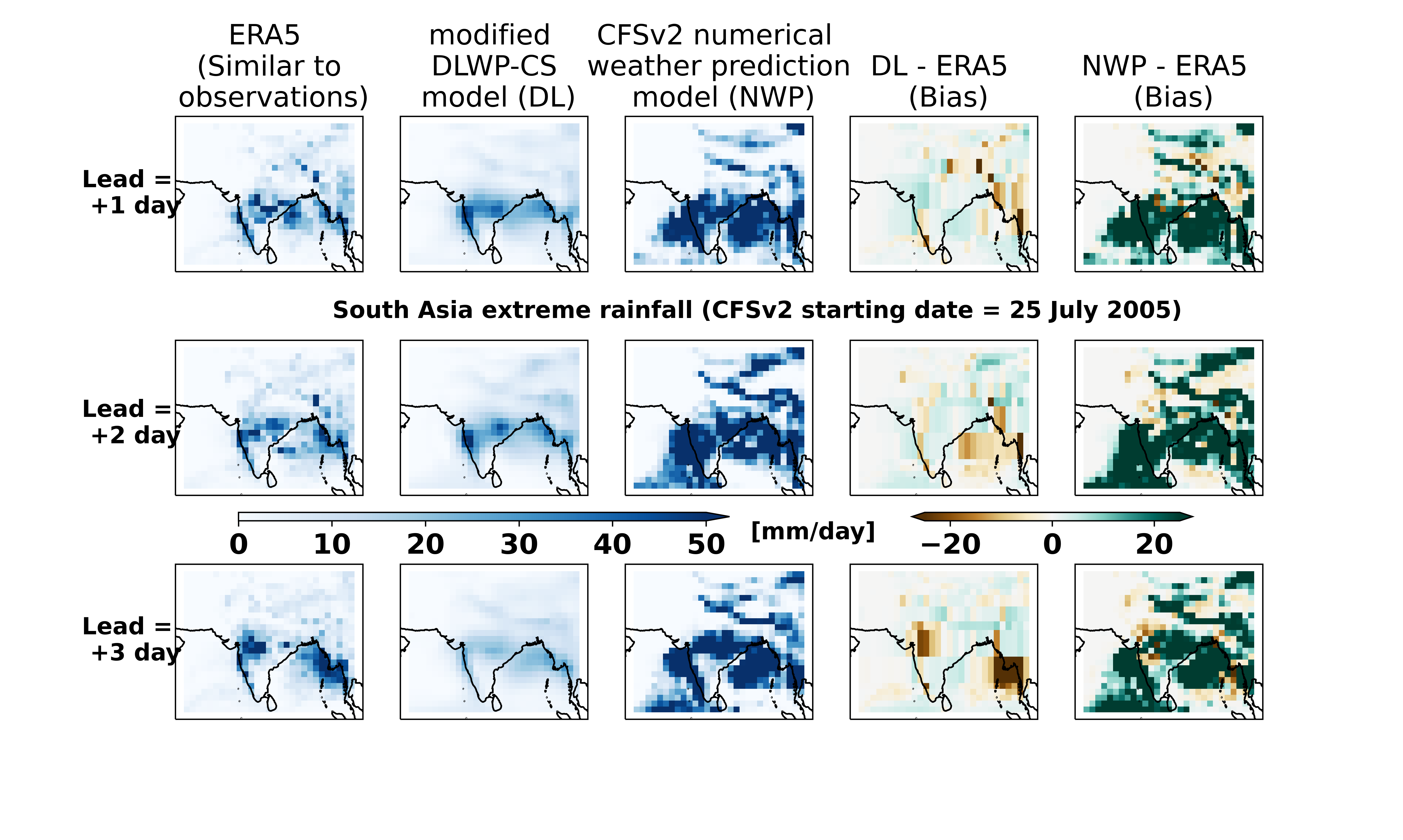}

  \caption{Absolute values corresponding to ERA5, deep learning augmented CFSv2 (DL) and CFSv2 are shown in the first three columns. Last two columns show the bias as difference between deep learning augmented CFSv2 and CFSv2 alone. The rows correspond to the different lead times, viz, 1, 2 and 3 days. The figure shows example for South Asia extreme precipitation in 2005.}
  \label{fig:india}
\end{figure*}

\begin{figure*}
  \centering
  \includegraphics[width=0.8\linewidth]{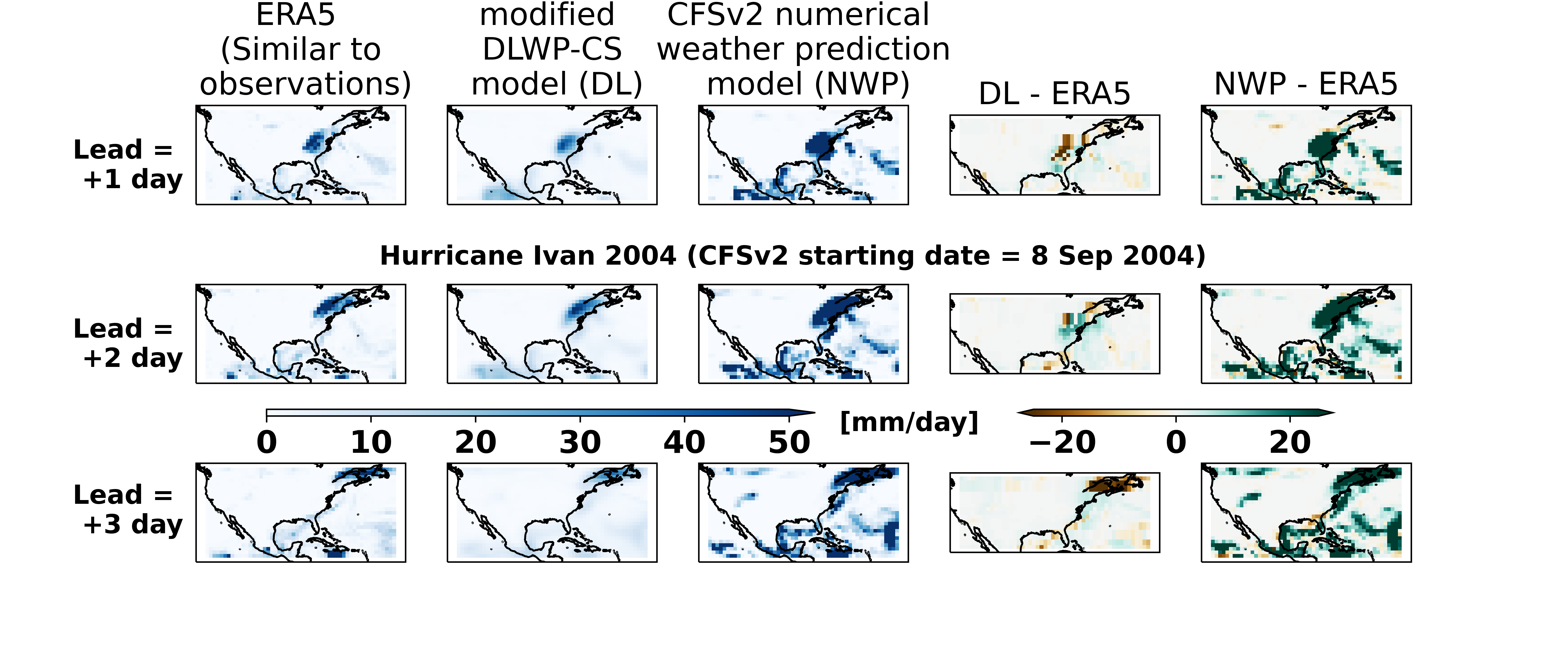}

  \caption{Same as figure \ref{fig:india} but for Hurricane Ivan in 2004.}
  \label{fig:ivan}
\end{figure*}

\begin{figure*}
  \centering
  \includegraphics[width=0.8\linewidth]{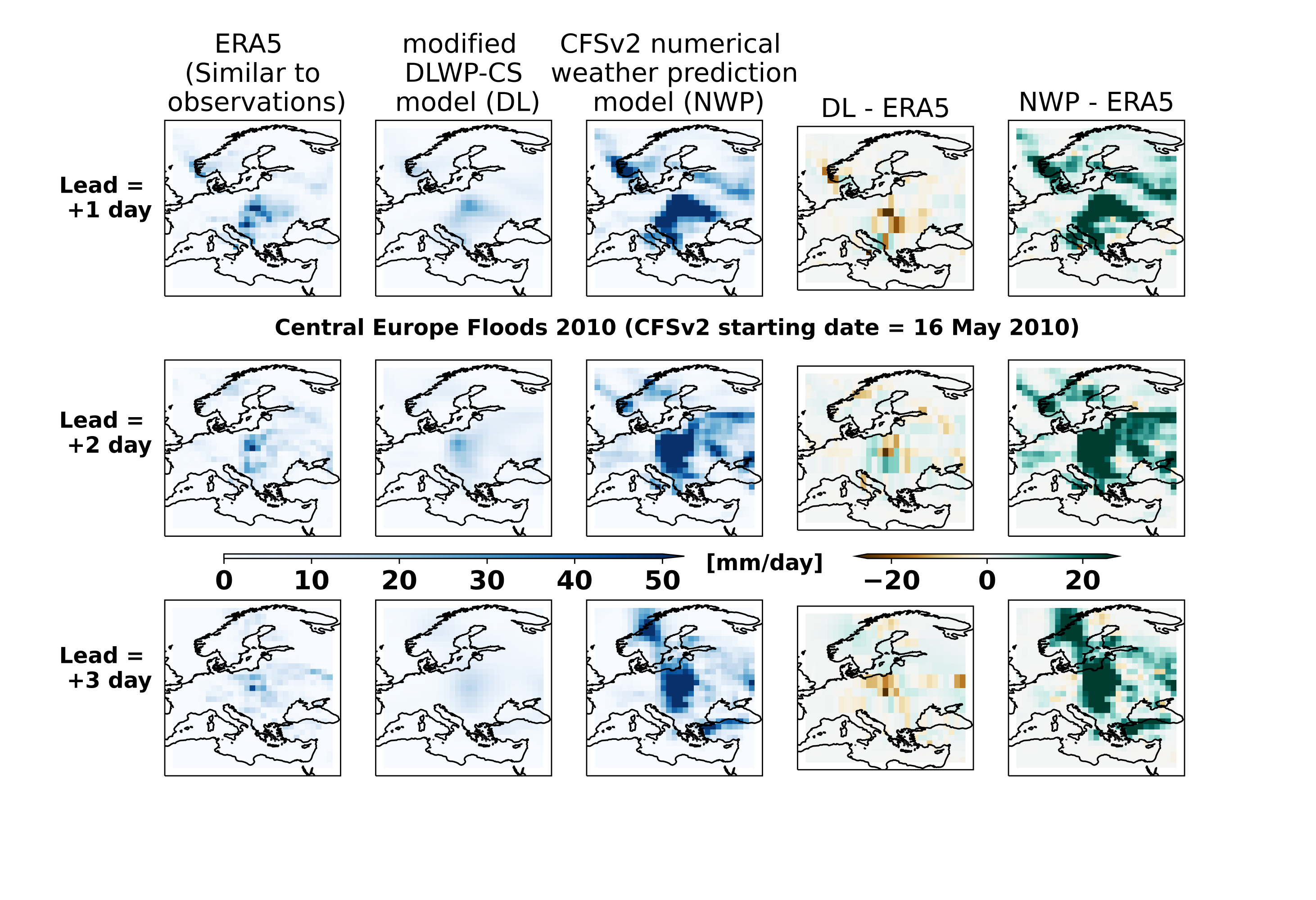}

  \caption{Same as figure \ref{fig:india} but for Europe flood in 2010.}
  \label{fig:europe}
\end{figure*}
\begin{figure*}
  \centering
  \includegraphics[width=0.8\linewidth]{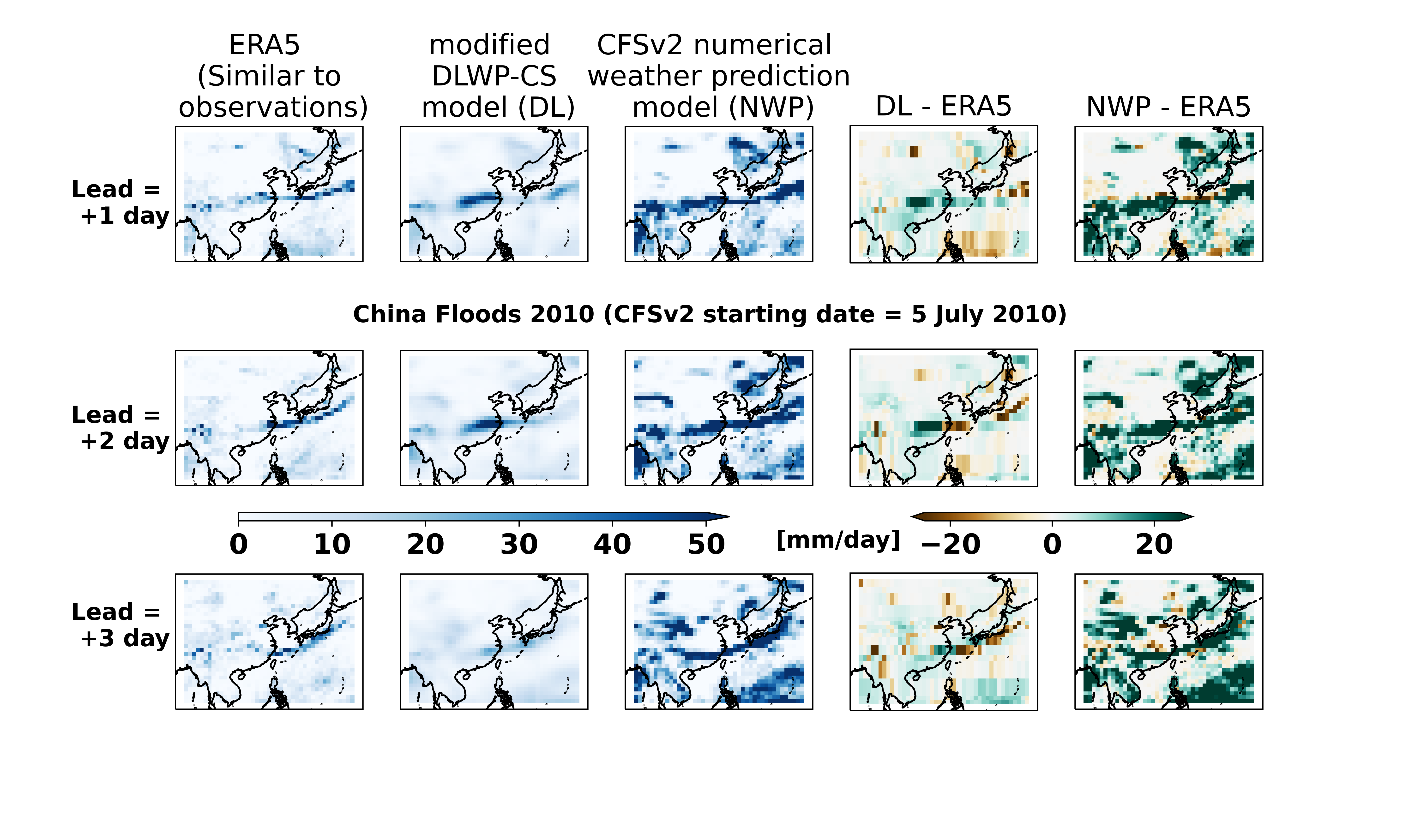}

  \caption{Same as figure \ref{fig:india} but for China floods in 2010.}
  \label{fig:china}
\end{figure*}

\begin{figure*}
  \centering
  \includegraphics[width=0.8\linewidth]{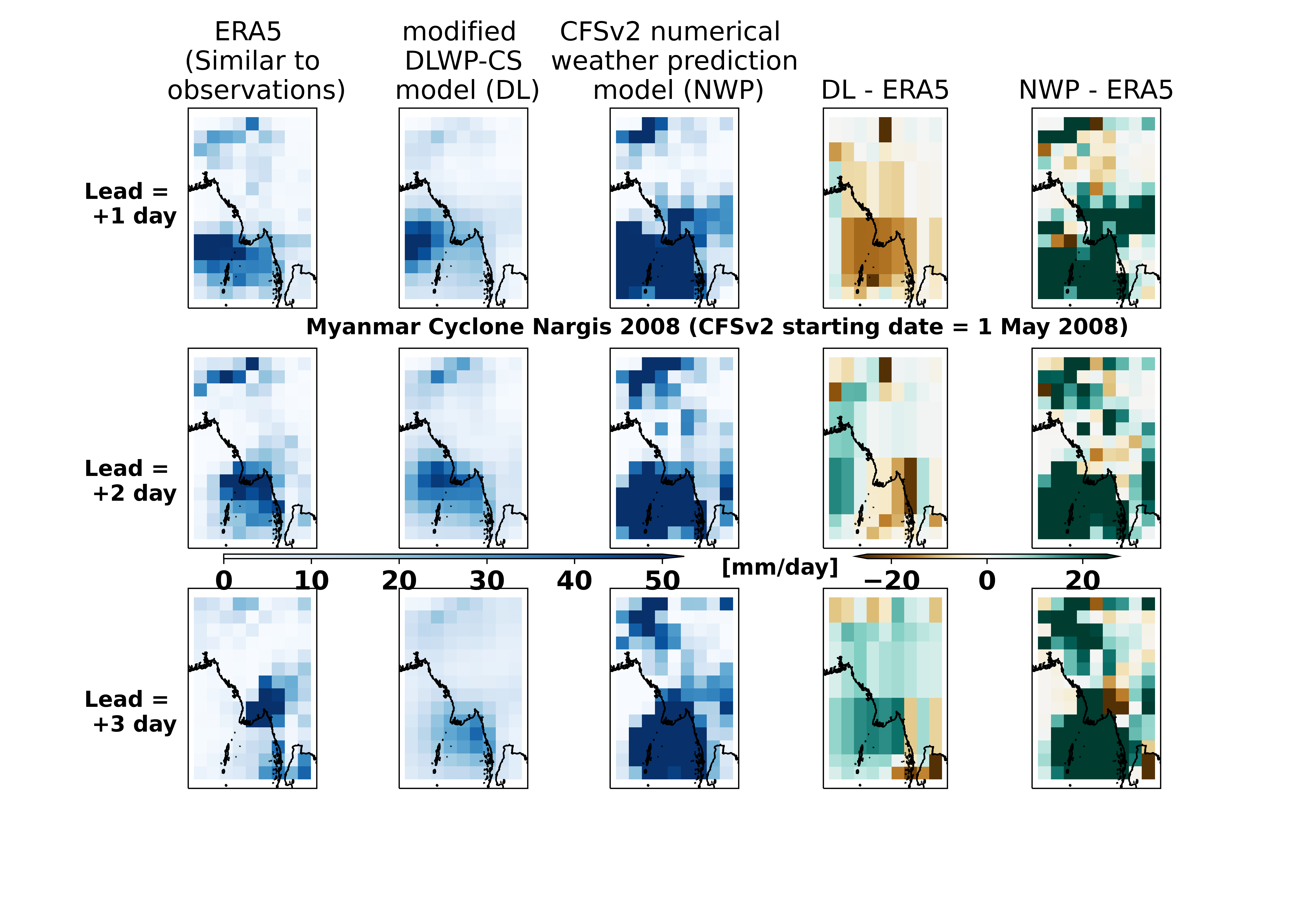}

  \caption{Same as figure \ref{fig:india} but for Myanmar cyclone Nargis that occurred in 2008}
  \label{fig:nargis}
\end{figure*}     

\begin{table*}
%  \centering
  \begin{tabular}{||c||c|c||c|c||c|c||}
\hline
\hline
\multicolumn{1}{|c|}{Events}
  & \multicolumn{2}{|c||}{Lead = 1 day (mm/day)} & \multicolumn{2}{|c||}{Lead = 2 day (mm/day)} & \multicolumn{2}{|c||}{Lead = 3 day (mm/day)}\\	
% \hline
\cline{2-7}
  &  DL - ERA5 & CFSv2 - ERA5 &  DL - ERA5 & CFSv2 - ERA5 &  DL - ERA5 & CFSv2 - ERA5\\
 \hline
 \hline

% China Floods & 05/07/2010 & \textbf{0.41} & 8.928 & \textbf{0.586} & 13.491 & \textbf{0.5}& 13.677\\

Hurricane Katrina  & \textbf{-0.345} & 8.839 & \textbf{0.453} & 12.18 & \textbf{-0.811} & 10.227\\

% Hurricane Ivan & 08/09/2004 & \textbf{-0.59} & 6.77 & \textbf{0.463} & 11.8 & \textbf{-0.58} & 12.476\\

Hurricane Ivan  & \textbf{-0.22} & 8.466 & \textbf{-0.036} & 13.48 & \textbf{-1.485} & 13.135\\

Cyclone Nargis  & \textbf{-5.37} & 21.151 & \textbf{-1.245} & 43.845 & \textbf{2.338} & 47.233\\

Europe Floods & \textbf{-0.2} & 6.654 & \textbf{-0.015} & 8.134 & \textbf{0.12} & 6.94\\
China Floods & \textbf{-0.17} & 11.233 & \textbf{0.465} & 18.903 & \textbf{-0.48} & 16.877\\
India flood  & \textbf{0.003} & 17.321 & \textbf{0.139} & 25.297 & \textbf{-0.749} & 20.259\\

% Europe Floods  & \textbf{-0.389} & 6.64 & \textbf{0.03} & 8.087 & \textbf{-0.873} & 4.453\\
\hline
\hline
  \end{tabular}
  \caption{Performance of the deep learning augmented numerical weather prediction system CFSv2 versus CFSv2 alone. The table shows regional bias/error in simulating various extreme precipitation events by the hybrid deep learning and CFSv2 system versus CFSv2 alone. The events occured as (i) Hurricane Katrina in 2005, (ii) Hurricane Ivan in 2004, (iii) Cyclone Nargis in 2008, (iv) Europe floods in 2010, (v) China flood in 2005 and (vi) India flood in 2005}
  \label{tab:example}
\end{table*}
\end{appendices}
\end{document}